\documentclass[preprint,review,12pt]{elsarticle}
\usepackage{graphicx}
\usepackage[utf8x]{inputenc}
\usepackage{color}
\usepackage{amsmath}
\usepackage{framed}

\begin{document}

\author[uzh]{Henrik Grundmann\corref{cor1}} 
\ead{grundmann@physik.uzh.ch}
\cortext[cor1]{Corresponding author. Tel.: ++41 44 635 5782; Fax: ++41 44 635 5704}

\author[uzh]{Andreas Schilling} 
\author[cam]{Casey A. Marjerrison} 
\author[had]{Hanna A. Dabkowska} 
\author[cam,had,bag]{Bruce D. Gaulin} 

\address[uzh]{Universit\"at Z\"urich, Physik-Institut,Winterthurerstr. 190, CH-8057 Zürich, {Switzerland}}
\address[cam]{{Department of Physics and Astronomy, McMaster University, Hamilton, Ontario, L8S 4M1, Canada}}
\address[had]{Brockhouse Institute for Materials Research, McMaster University, Hamilton, Ontario, L8S 4M1, Canada}
\address[bag]{{Canadian Institute for Advanced Research, 180 Dundas Street W, Toronto, Ontario, M5G 1Z8, Canada}}

\title{Structure and magnetic interaction{s} in the solid solution Ba\(_{3−x}\)Sr\(_x\)Cr\(_2\)O\(_8\)}

\begin{abstract}
 \noindent Solid solutions of the magnetic insulators Ba\(_3\)Cr\(_2\)O\(_8\) and Sr\(_3\)Cr\(_2\)O\(_8\) (Ba\(_{3-x}\)Sr\(_x\)Cr\(_2\)O\(_8\)) have been prepared in polycrystalline form for the first time. Single crystalline material was obtained using a mirror image floating zone technique. X-ray diffraction data taken at room temperature indicate that the space group of Ba\(_{3-x}\)Sr\(_x\)Cr\(_2\)O\(_8\) remains unchanged for all values of \textit{x}, while the cell parameters depend on the chemical composition, as expected. Magnetization data, measured from 300\,K down to 2\,K, suggest that the interaction constant \(J_\mathrm{d}\) within the Cr\(^{5+}\) dimers varies in a peculiar way as a function of \textit{x}, starting at \(J_\mathrm{d}=25\)\,K for \(x=0\), then first slightly dropping to \(J_\mathrm{d}=18\)\,K for \(x\approx0.75\), before reaching  \(J_\mathrm{d}=62\)\,K for \(x=3\).
\end{abstract}
\maketitle
\section{Introduction}
In the last years, many spin dimer systems have been shown to exhibit a field induced phase transition of the magnetic subsystem \cite{shiramura_tlcucl3,jaime_bacusi2o6} below a critical temperature and above a certain critical magnetic field \(H_\text{c}\). This phenomenon has been interpreted in terms of a Bose-Einstein condensation {(BEC)} of the magnetic quasiparticles (triplons) in connection with the collective dimer states\cite{nikuni_tlcucl3_magnetization}.

The materials Ba\(_3\)Cr\(_2\)O\(_8\) and Sr\(_3\)Cr\(_2\)O\(_8\) are examples for such spin dimer systems, and they have been considered as possible candidates for a triplon BEC \cite{buschbaum_ba3cr2o8,buschbaum_sr3cr2o8,nakajima_ba3cr2o8,singh_sr3cr2o8,aczel_ba3cr2o8_triplonbec,aczel_ba3cr2o8_jcg,aczel_sr3cr2o8,Kamenskyi_ba3cr2o8}. Recently, these two compounds have been proposed as candidates for a new experimental scheme to probe the macroscopic phase coherence that should be found in all Bose-Einstein condensates \cite{schilling_josephson}.
It has been argued that {compounds in} a solid solution of Ba\(_3\)Cr\(_2\)O\(_8\) and Sr\(_3\)Cr\(_2\)O\(_8\) might have different critical field{s} than the mother compounds, but with otherwise similar physical properties. The chemical potential \({\mu}\) for the triplon quasiparticles in such {materials} is controlled by the external magnetic field \({H}\) through \(\mu=g\mu_\mathrm{B}\mu_0(H-H_\mathrm{c})\) \cite{giamarchi_magneticisolator-bec} with \(\mu_\mathrm{B}\) the Bohr magneton and \(g\) the Landé factor. A change in the chemical potential stemming from, e.g., a changed stoichiometry leading to slightly differing values for \(H_\text{c}\) between two compounds, might induce Josephson effects in a device of two of such  crystals coupled together\cite{schilling_josephson}. We therefore systematically studied the system Ba\(_{3-x}\)Sr\(_x\)Cr\(_2\)O\(_8\) to obtain information about its physical properties, in particular about the magnetic interaction constants that determine the value of the critical fields \(H_\mathrm{c}\).
\section{Experimental details}
\begin{figure}
 \centering
 \includegraphics[width=.4\textwidth]{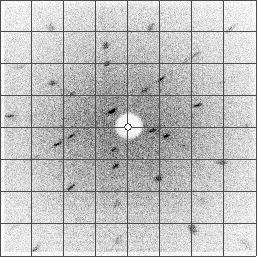}
 \caption{X-ray Laue diffractogram for the grown boule of Ba\(_{2.9}\)Sr\(_{0.1}\)Cr\(_2\)O\(_8\). The beam direction was close to the \textit{c}-axis. The hexagonal symmetry is clearly visible.}
 \label{fig:ba29laue}
\end{figure}

The samples were first synthesized as polycrystalline powders using standard solid-state reaction schemes. We used three different methods to prepare the polycrystalline material.
First, Ba(NO\(_3\))\(_2\), Sr(NO\(_3\))\(_2\) and Cr(NO\(_3\))\(_2\cdot\)9H\(_2\)O were used as reactants. The powders were mixed according to 
\begin{align*}
     \text{(3-x)Ba(NO}_3\text{)}_2 + \text{xSr(NO}_3\text{)}_2 + \text{2Cr(NO}_3\text{)}_2\cdot\text{9H}_2\text{O}\rightarrow&\text{Ba}_\text{3-x}\text{Sr}_\text{x}\text{Cr}_2\text{O}_8+\\
&+\text{10NO}_2+\text{O}_2+9\text{H}_2\text{O},
\end{align*}
dissolved in water and heated afterwards while {continuously} stirred to keep the mixture homogeneous. After evaporating the water, the remaining powder was ground and heated under flowing argon at 915\,\(^{\circ}\)C for 24\,h to remove any excess water and NO\(_x\). The resulting oxide powder was ground again, mechanically pressed into pellets and sintered at 1100\,\(^\circ\)C for 48\, h under flowing Ar. For \(x=0.1\) and \(x=1.5\), rods were hydrostatically pressed and annealed under the same conditions. The obtained polycrystalline material was black.
An optical floating zone method was used to grow crystals from those rods \cite{dabko_floatingzone}. Two of the resulting polycrystalline rods were then mounted in a \textit{Crystal Systems Inc.} image furnace using Pt wire, a short one (about 3\,cm) as a seed rod, and a longer one (about 10\,cm)  as a feed rod.
The crystal growth was performed in two stages, both in a high-purity environment of flowing Ar. The first stage was performed relatively fast (growth rates between 21\,mm/h and 25\,mm/h) to produce a premelted rod with a high density. For the second stage, the premelted rod was used as a feed rod with the remnants of the former feed rod as a seed. The growth rate was relatively low (between 2\,mm/h and 5\,mm/h), but was varied during the growth to maintain the zone stability.
The resulting boule consisted of several, well oriented grains (see Fig. \ref{fig:ba29laue} for the X-ray Laue-diffractogram).

While we obtained crystalline samples for \(x={0.1}\), attempts with \(x=1.5\) have not yet resulted in crystalline material.

Additionally, polycrystalline samples of several stoichiometries were prepared according to
\begin{equation*}
     \text{(3-x)BaCO}_3 + \text{xSrCO}_3 + \text{Cr}_2\text{O}_3 + \text{O}_2\rightarrow\text{Ba}_\text{3-x}\text{Sr}_\text{x}\text{Cr}_2\text{O}_8+3\text{CO}_2
\end{equation*}

and 

\begin{equation*}
     2\text{(3-x)BaCO}_3 + 2\text{xSrCO}_3 + 4\text{Cr}\text{O}_3\rightarrow2\text{Ba}_\text{3-x}\text{Sr}_\text{x}\text{Cr}_2\text{O}_8+6\text{CO}_2 + \text{O}_2{.}
\end{equation*}

The respective carbonates and chromium oxides were mixed, ground and heated under flowing argon at 1000\,°C for 24\,h. The resulting material was ground again and sealed in an alumina crucible inside a quartz tube under vacuum to prevent a reaction between sample and tube. The tube was heated up to 1200\,°C for 24\,h and quenched {into} water. This last step was repeated several times, with intermediate grindings.
Each of the three preparation methods yielded homogeneous black polycrystalline material and we did not find relevant differences {between the respective products} for any value of \textit{x} between 0 and 3. However, we noted that grinding the material again thoroughly in air or dry He-atmosphere {led} to a slightly green color of the ground powder for every value of \textit{x}.

The results shown below were obtained from the samples prepared by the first method {described} above, and from single crystalline material (for \(\text{Ba}_\text{2.9}\text{Sr}_\text{0.1}\text{Cr}_2\text{O}_8\)).

\section{Crystallographic structure}
\begin{figure}[h]
 \centering
 \includegraphics{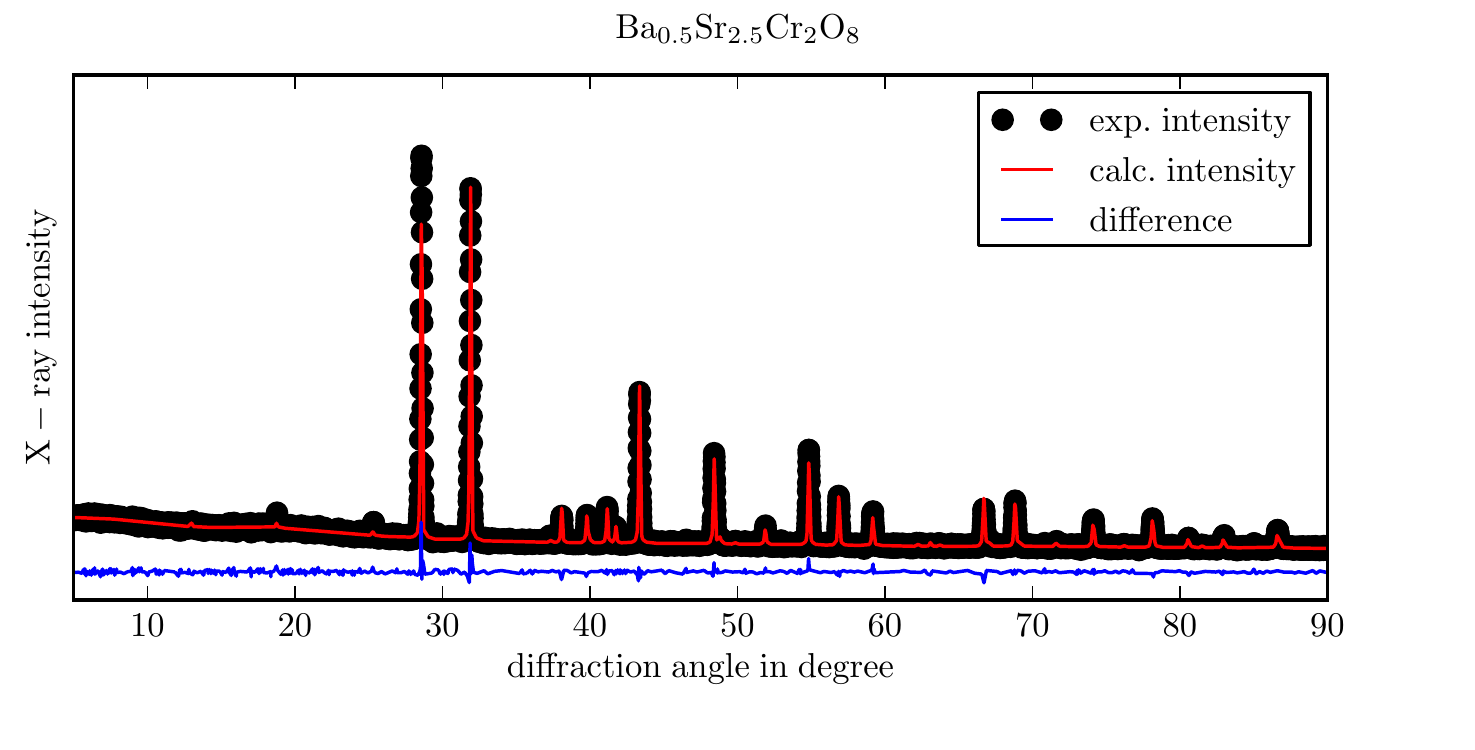}
 \caption{Experimental and calculated X-ray diffraction intensities for Ba\(_{0.5}\)Sr\(_{2.5}\)Cr\(_2\)O\(_8\)}
 \label{fig:xrayba0p5sr2p5cr2o8}
\end{figure}

\begin{figure}
 \centering
 \includegraphics{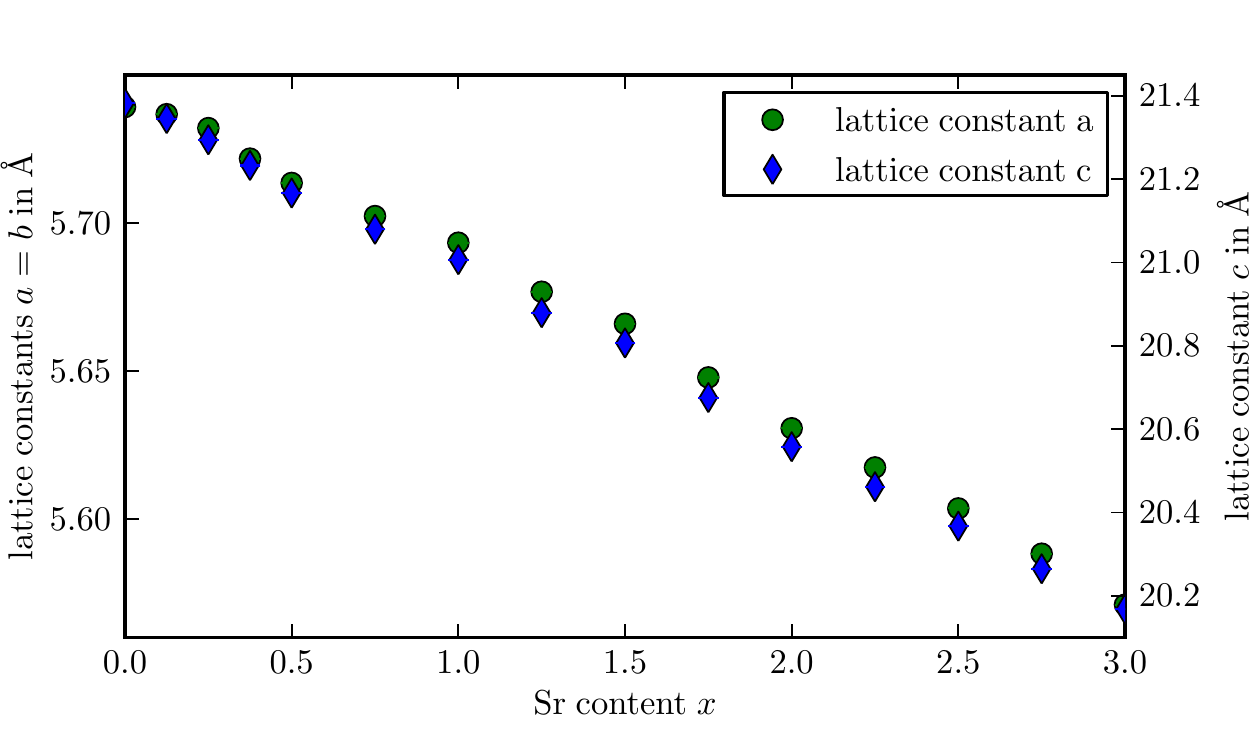}
 \caption{Lattice constants \(a=b\) and \(c\) of Ba\(_{3-x}\)Sr\(_x\)Cr\(_2\)O\(_8\) as obtained from X-ray diffraction data, as functions of the strontium content \textit{x}.}
 \label{fig:latticeconstants}
\end{figure}

\begin{figure}
 \centering
 \includegraphics{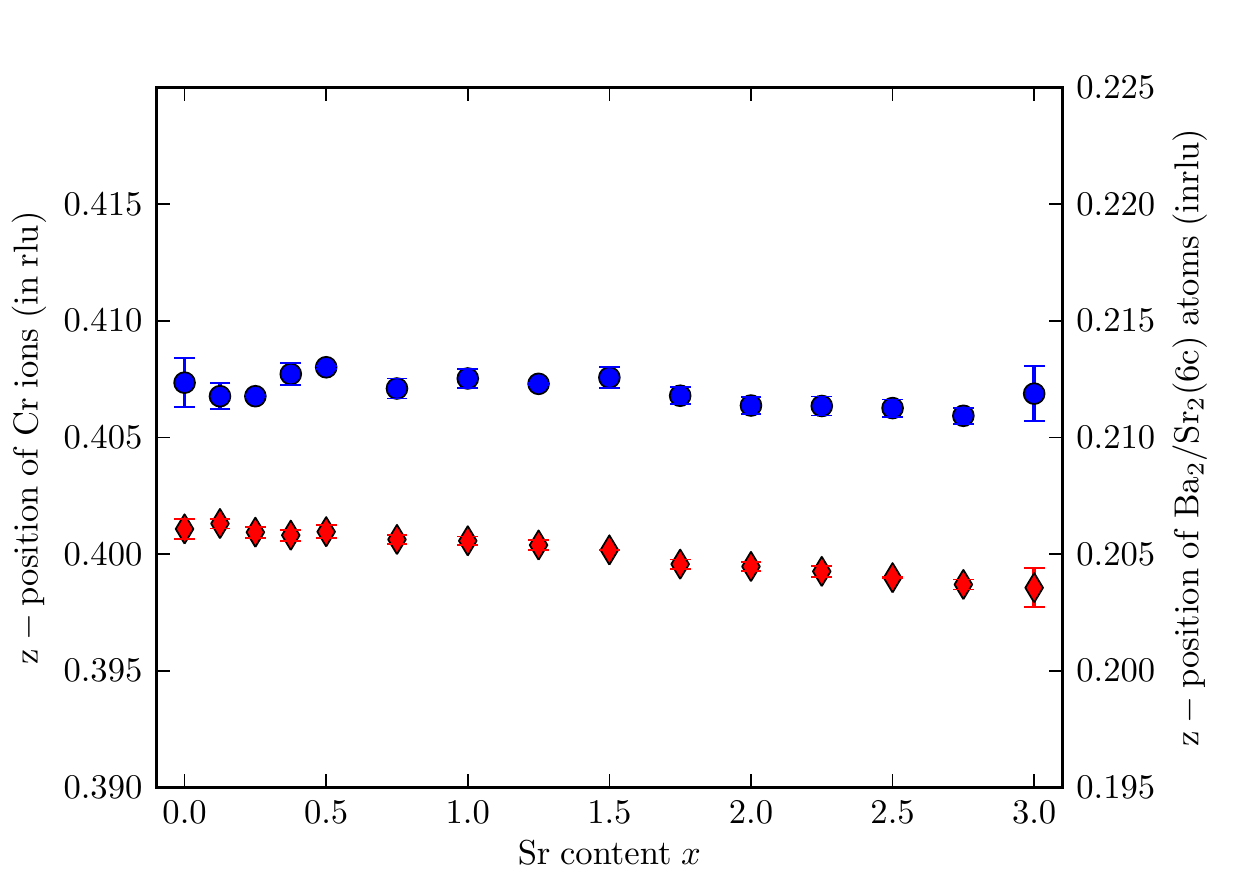}
 \caption{\textit{z}-value of the atomic position for the Cr (circles) and Ba\(_2\)/Sr\(_2\)(nomenclature according to \cite{kofu_ba3cr2o8}) (diamonds) ions as a function of the strontium content \textit{x}. The \textit{x}- and \textit{y}-values where kept at zero.}
 \label{fig:crpos}
\end{figure}

\begin{figure}
 \centering
 \includegraphics{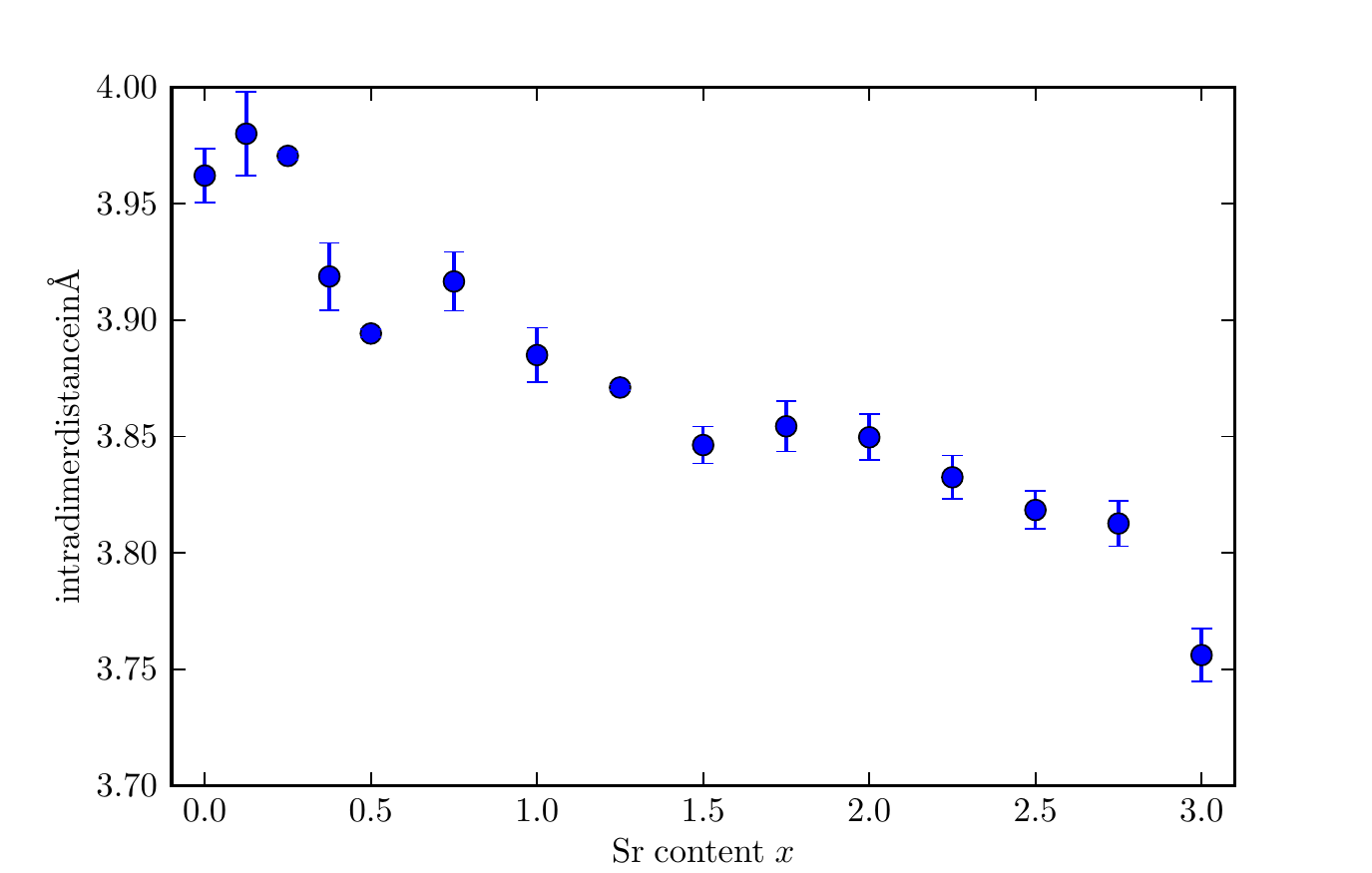}
 \caption{Distance between the Cr ion constituting one dimer as a function of the strontium content \textit{x}.}
 \label{fig:dimerdistance}
\end{figure}

{In order to perform} structural analysis, parts of the boule as well as corresponding polycrystalline samples with \(0{\leq}x{\leq}3\) were ground and examined using Cu\(_{\text{K}\alpha}\) {X-ray} radiation. The resulting patterns were analyzed using the Rietveld method (\textit{FullProf suite}). As expected, no difference was found between ground single crystals and polycrystalline material.
All of the observed reflections of the mixed crystals are in good agreement with the space group R\(\overline{3}\)m (see Fig.  \ref{fig:xrayba0p5sr2p5cr2o8}), which is the same as that of the mother compounds with \(x=0\) and \(x=3\). As scattering coefficients for Cr\(^{5+}\) were not available, we used the known values for Cr\(^{3+}\).
{The absolute values of the cell parameters depend linearly on the strontium content \textit{x}, in agreement with Vegard's law (see Fig. \ref{fig:latticeconstants}). In Fig. \ref{fig:dimerdistance}, we have plotted the distance between two Cr ions forming a single dimer. As expected, this intradimer distance also decreases smoothly as a function of \textit{x}. We did not find any relevant systematic change in the relative atomic positions as a function of \textit{x}, although it should be noted that the sensitivity on the oxygen position is not high enough to exclude such a shift. In Fig. \ref{fig:crpos}, we show the \textit{z}-value of the atomic position for the Cr and Ba/Sr ions while the respective \textit{x} and \textit{y} coordinates were assumed to be zero. All other coordinates where kept zero for all \textit{x}.}

\section{Magnetic properties}
\begin{figure}
 \centering
 \includegraphics{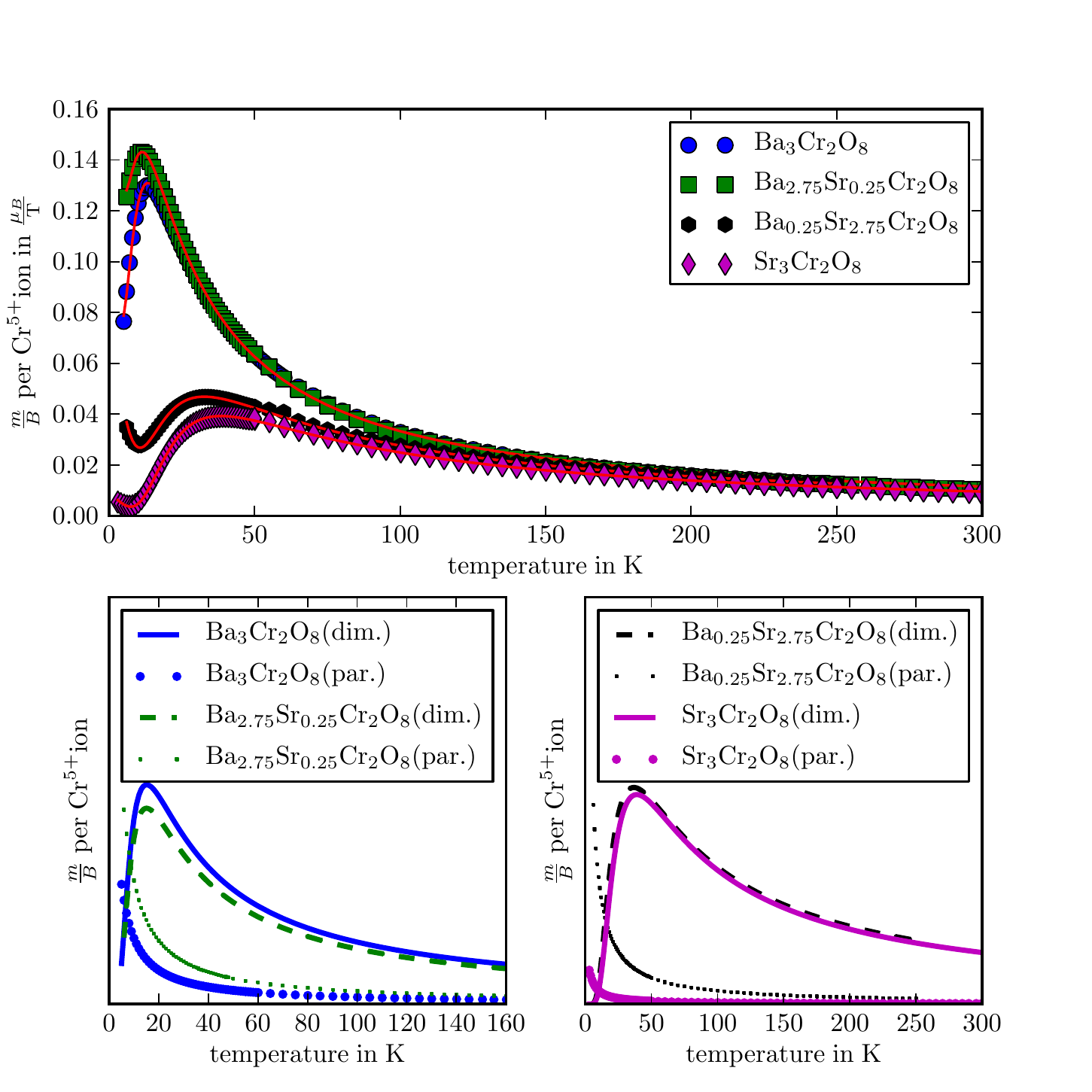}
 \caption{Upper panel: magnetization data \({\frac{m}{B}(T)}\) for four different stoichiometries of Ba\(_{3-x}\)Sr\(_x\)Cr\(_2\)O\(_8\) in \(\mu_0H=1\)\,T. The solid lines correspond to fits according to a sum of Eqs. \ref{eqn:dimer} and \ref{eqn:paramag}. The lower panels show the respective extracted paramagnetic (par.) and dimer (dim.) contributions to the total magnetization.}
 \label{fig:fiterlaeuterung}
\end{figure}

The magnetic properties of the polycrystalline samples were analyzed using a commercial SQUID magnetometer (\textit{Quantum Design Inc.}). The magnetization was measured for temperatures between 5\,K and 300\,K and in magnetic fields of \(\mu_0H=1\,\text{T}\) and \(\mu_0H=5\,\text{T}\). For compositions with \(x<0.2\) and \(x>2.8\), the magnetization shows a pronounced maximum at low temperatures. The experimental data can be well described with the Bleaney-Bowers formula for interacting dimers,
\begin{equation}                                                                                                                                                                                                                                                                                                                                                                                                                                                                               
M_d(T)=\frac{n_\mathrm{d}g^2\mu_\mathrm{B}B_\text{ext}}{J_e+k_BT\left(3+e^{\frac{J_\mathrm{d}}{k_BT}}\right)} \label{eqn:dimer},                                                                                                                                                                                                                                                                                                                                                                                                                                                                         \end{equation}
with \(g=1.94\) which is only slightly anisotropic in these systems\cite{kofu_ba3cr2o8,quinteracastro_sr3cr2o8}, \(n_\mathrm{d}\) the density of the coupled ions and \(J_e\) and \(J_\mathrm{d}\) the inter- and intradimer coupling constant{s}, respectively. For intermediate values of \textit{x} around 1.5, a paramagnetic background becomes more and more relevant so that a Brillouin term had to be included, 
\begin{equation}M_p(T)=n_Pg\mu_\mathrm{B}\frac12\left(2\coth\left(\frac{g\mu_\mathrm{B}B_\text{ext}}{k_BT}\right)-\coth\left(\frac{g\mu_\mathrm{B}B_\text{ext}}{2k_BT}\right)\right) \label{eqn:paramag},
\end{equation}
where \(n_P\) denotes the density of the corresponding uncoupled ions. 
The experimental data are reasonably well described by the sum of {the} above terms (see Fig. \ref{fig:fiterlaeuterung}). However, as the fit is not very sensitive to the interdimer interaction constant \(J_e\) (especially in the presence of a strong paramagnetic background) we could not extract reliable values for it.

\begin{figure}
 \centering
 \includegraphics{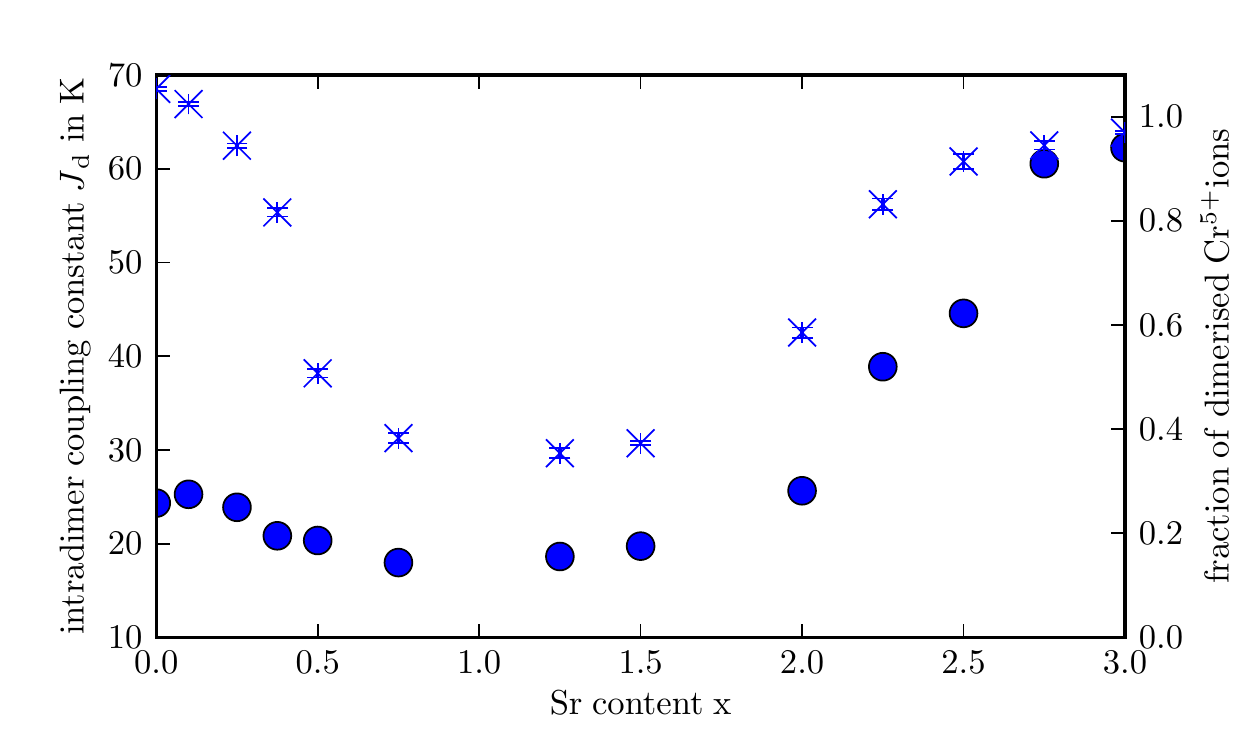}
 \caption{Estimated intradimer interaction constant \(J_\text{d}\) in Ba\(_{3-x}\)Sr\(_x\)Cr\(_2\)O\(_8\) (filled circles) and the fraction of dimerized {Cr}\(^{5+}\)-ions (crosses, prefactor \(n_\mathrm{d}\) from Eq. \ref{eqn:dimer}), as a function of the strontium content \textit{x}.}
 \label{fig:j_intra}
\end{figure}

\begin{figure}[h!]
 \centering
 \includegraphics{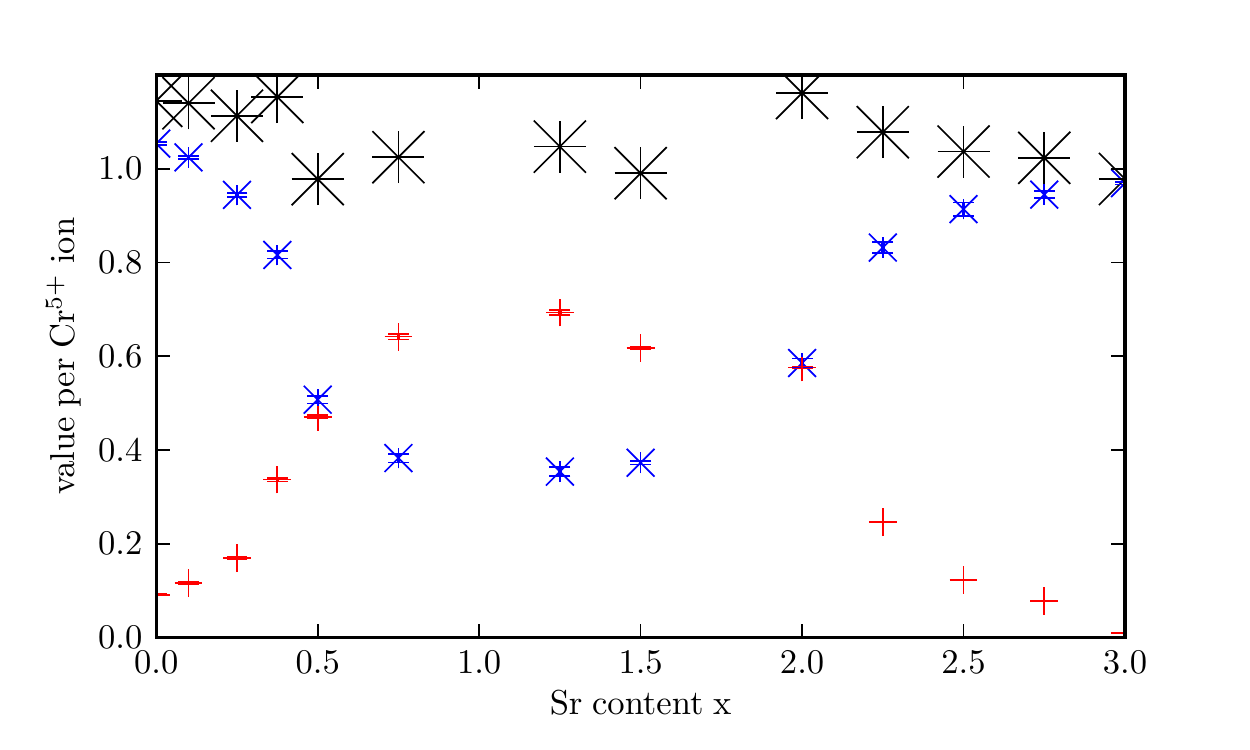}
 \caption{Estimated fraction \(n_\mathrm{d}\) of dimerized (crosses) and free Cr\(^{5+}\)-ions \({n_\mathrm{p}}\) (plus signs) from Eqs. \ref{eqn:dimer} and \ref{eqn:paramag} and the sum \(n_\mathrm{d}+n_\mathrm{p}\) (stars), as functions of the strontium content \textit{x}.}
 \label{fig:background}
\end{figure}

The thus obtained intradimer interaction constant \(J_\mathrm{d}\) strongly depends on the stoichiometry, as shown in Fig. \ref{fig:j_intra}. Surprisingly, it first decreases with increasing strontium content up to \(x\approx1\), before increasing again up to the value for pure Sr\(_3\)Cr\(_2\)O\(_8\). As the relative atomic positions do not change with \textit{x} (see Fig. \ref{fig:crpos}), we attribute the overall increase of the interaction constant to a decrease of the absolute intradimer distance of the Cr\(^{5+}\)-ions from pure  Ba\(_3\)Cr\(_2\)O\(_8\) to pure Sr\(_3\)Cr\(_2\)O\(_8\) {(see Fig. \ref{fig:dimerdistance})}. As a consequence, the variation of \(J_\mathrm{d}\) with \textit{x} should lead to an associated change in the critical field \(H_\mathrm{c}\), which we will investigate in a separate report.

\begin{figure}[h]
 \centering
 \includegraphics{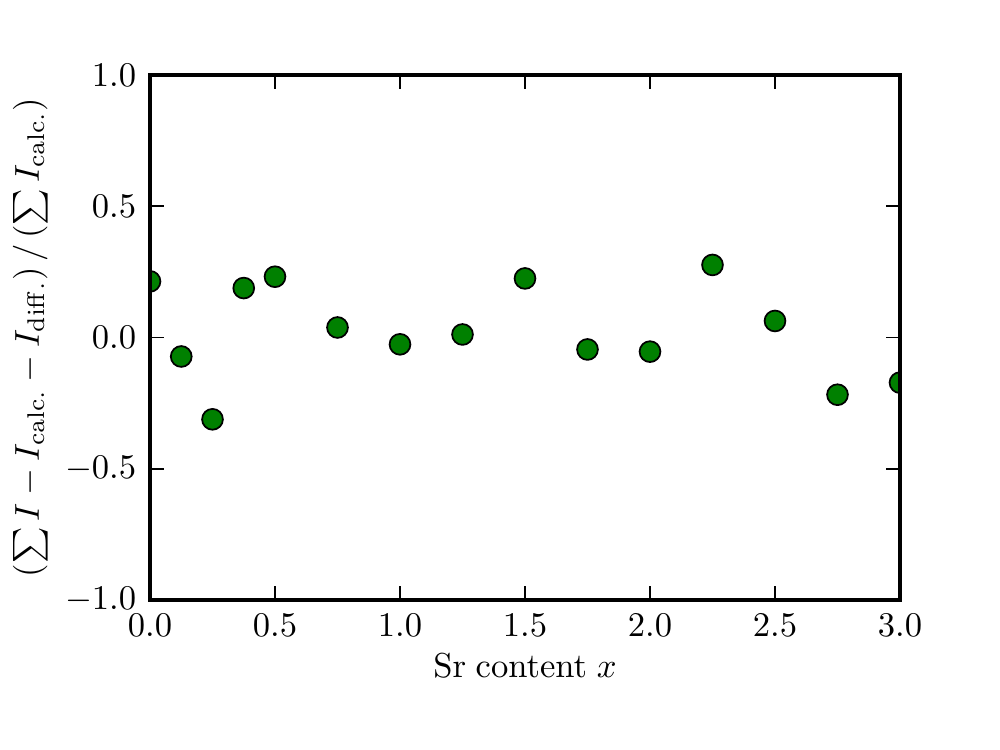}
 \caption{Integrated scattering intensity of potential impurities.}
 \label{fig:fremdcounts}
\end{figure}

In Fig. \ref{fig:j_intra} we also show the fraction \(n_\mathrm{d}\) of dimerized Cr\(^{5+}\) ions and we compare it in Fig. \ref{fig:background} with the density \(n_\mathrm{p}\) of uncoupled Cr\(^{5+}\) ions, both treated as independent fitting parameters. The paramagnetic background for intermediate \textit{x} is substantial, but its microscopic origin is not clear up to now. The sum of the prefactors \(n_\mathrm{d}\) and \(n_\mathrm{p}\) is reasonably close to 1, which supports the validity of our fitting procedure. An increase of the preparation temperature to 1250\,°C as well as repeated grinding and annealing of the polycrystalline samples did not reduce this background contribution substantially, but our corresponding X-ray diffraction data do, most interestingly, not suggest the appearance of possible impurity phases. {In Fig. \ref{fig:fremdcounts} we show the part of the total scattering intensity that is neither associated with the expected diffraction peaks of Ba\(_{3-x}\)Sr\(_x\)Cr\(_2\)O\(_8\) (\(I_\text{calc.}\)) nor with the diffuse background (\(I_\text{diff.}\)), normalized to the total integrated intensity without the diffuse background. Within the accuracy of this procedure, diffraction peaks stemming from unwanted impurity phases are virtually absent for all values of \textit{x}}. We therefore believe that this magnetic background is intrinsic to the polycrystalline system Ba\(_{3-x}\)Sr\(_x\)Cr\(_2\)O\(_8\).

\section{Conclusion}
We have for the first time synthesized mixed crystals of the magnetic insulators Ba\(_{3-x}\)Sr\(_x\)Cr\(_2\)O\(_8\). The structure of the solid solution series remains unchanged, but with linearly decreasing cell parameters for increasing Sr-content \textit{x}. The magnetic behavior indicates the presence of a dimerized spin system for all values of \textit{x}, with an interaction constant \(J_\mathrm{d}\) that strongly depends on the composition and surprisingly falls below the value of pure Ba\(_3\)Cr\(_2\)O\(_8\) in certain range of \textit{x}. For intermediate values (\(x\approx0.5\,...\,2.0\)), a paramagnetic background appears that does not seem to be related to impurity phases and is therefore probably intrinsic to Ba\(_{3-x}\)Sr\(_x\)Cr\(_2\)O\(_8\).

\section{Acknowledgements}
We would like to thank Dr. M. Medarde Barragan and Dr. A. Krzton-Maziopa for valuable discussions. We also would like to thank T. Munsie and Antoni B. Dabkowski for their help in the crystal growth and characterization. This work was supported by the Swiss National Science Foundation Grants No. 21-126411 and 21-140465.

\bibliographystyle{elsarticle-num}
\bibliography{ba3-xsrxcr2o8_strukturpaper}

\end{document}